\documentclass[article,twocolumn, superscriptaddress, 10pt, nofootinbib]{revtex4-1}
\usepackage[sort&compress]{natbib}   
\usepackage{amsmath,amssymb,amsfonts}       
\usepackage{graphicx}                       

\usepackage[dvipsnames]{xcolor}                        
\usepackage[colorlinks=true]{hyperref}    
\usepackage[UKenglish]{babel}
\usepackage[UKenglish]{isodate}

\usepackage[utf8]{inputenc}
\usepackage[T1]{fontenc} 
\usepackage{braket} 
\usepackage[normalem]{ulem} 

\usepackage{multirow}
\usepackage{braket}
\usepackage{ulem}
\usepackage{enumitem}
\usepackage{siunitx}

\begin{document}
	\title{Three-dimensional electrical control of the excitonic fine structure  for a quantum dot in a  cavity}

\author{H. Ollivier}
\affiliation{Universit\'e Paris-Saclay, CNRS, Centre de Nanosciences et de Nanotechnologies (C2N), 10 Boulevard Thomas Gobert, 91120 Palaiseau, France}
\author{P. Priya}
\affiliation{Universit\'e Paris-Saclay, CNRS, Centre de Nanosciences et de Nanotechnologies (C2N), 10 Boulevard Thomas Gobert, 91120 Palaiseau, France}
\author{A. Harouri}
\affiliation{Universit\'e Paris-Saclay, CNRS, Centre de Nanosciences et de Nanotechnologies (C2N), 10 Boulevard Thomas Gobert, 91120 Palaiseau, France}
\author{I. Sagnes}
\affiliation{Universit\'e Paris-Saclay, CNRS, Centre de Nanosciences et de Nanotechnologies (C2N), 10 Boulevard Thomas Gobert, 91120 Palaiseau, France}
\author{A. Lemaître}
\affiliation{Universit\'e Paris-Saclay, CNRS, Centre de Nanosciences et de Nanotechnologies (C2N), 10 Boulevard Thomas Gobert, 91120 Palaiseau, France}
\author{O. Krebs}
\affiliation{Universit\'e Paris-Saclay, CNRS, Centre de Nanosciences et de Nanotechnologies (C2N), 10 Boulevard Thomas Gobert, 91120 Palaiseau, France}
\author{L. Lanco}
\affiliation{Universit\'e Paris-Saclay, CNRS, Centre de Nanosciences et de Nanotechnologies (C2N), 10 Boulevard Thomas Gobert, 91120 Palaiseau, France}
\affiliation{Université de Paris, Centre de Nanosciences et de Nanotechnologies (C2N), 91120 Palaiseau, France}
\author{N. D. Lanzillotti-Kimura}
\affiliation{Universit\'e Paris-Saclay, CNRS, Centre de Nanosciences et de Nanotechnologies (C2N), 10 Boulevard Thomas Gobert, 91120 Palaiseau, France}
\author{M. Esmann}
\email{m.esmann@uni-oldenburg.de}
\affiliation{Universit\'e Paris-Saclay, CNRS, Centre de Nanosciences et de Nanotechnologies (C2N), 10 Boulevard Thomas Gobert, 91120 Palaiseau, France}
\affiliation{current affiliation: Institute of Physics, Carl von Ossietzky University, 26129 Oldenburg, Germany}

\author{P. Senellart}
\email{pascale.senellart-mardon@c2n.upsaclay.fr}
\affiliation{Universit\'e Paris-Saclay, CNRS, Centre de Nanosciences et de Nanotechnologies (C2N), 10 Boulevard Thomas Gobert, 91120 Palaiseau, France}

	\date{\today}
	
	\begin{abstract}
The excitonic fine structure plays a key role for the quantum light generated by semiconductor quantum dots, both for entangled photon pairs  and  single photons. Controlling  the excitonic fine structure has been demonstrated using electric, magnetic, or strain fields, but not for quantum dots in optical cavities, a key requirement to obtain high source efficiency and near-unity photon indistinguishability. Here, we demonstrate the control of the fine structure splitting for quantum dots embedded in micropillar cavities. We propose a scheme based on remote electrical contacts connected to the pillar cavity through narrow ridges. Numerical simulations show that such a geometry allows for a three-dimensional control of the electrical field. We experimentally demonstrate tuning and reproducible canceling of the fine structure, a  crucial step for the reproducibility of quantum light source technology. 
	\end{abstract}
	\maketitle

\begin{figure}
    \includegraphics{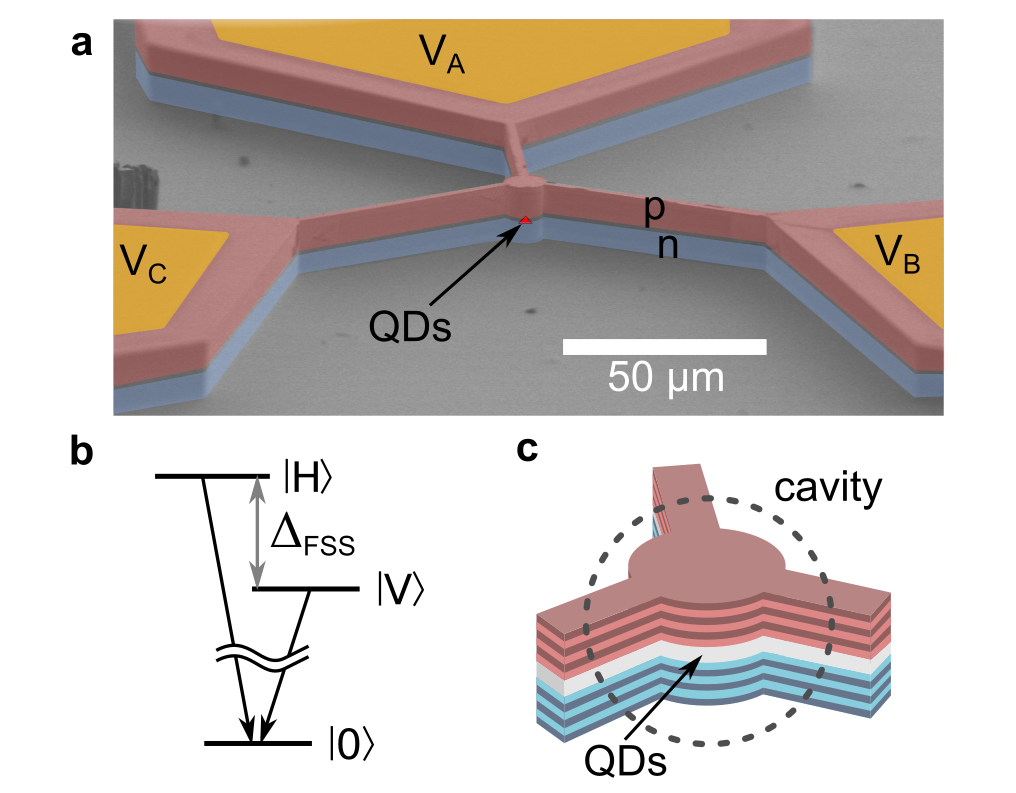}
    \caption{{(a)  SEM image of a 10~\textmu m diameter micropillar cavity with QDs embedded at its center. Ridges connect the pillar to three  electrical contacts defined via gold pads. The image has been artificially colored to evidence the different components of the structure (b) Schematic  of the exciton energy levels. (c) Schematic zoom of the pillar  structure highlighting the top and bottom cavity mirrors defining the cavity in the vertical direction. The QD layer is vertically centered in the cavity.}  }
    \label{fig1}
\end{figure}

Semiconductor quantum dots (QDs) have emerged as excellent emitters of  single photons~\cite{senellart_high-performance_2017} and entangled photon pairs~\cite{review-entangled-1,review-entangled-2} with exciting prospects for both optical quantum networks and processors. The neutral exciton state in an epitaxially grown QD  exhibits a fine structure splitting (FSS) arising from a reduced symmetry of the nanostructure as well as valence band mixing (see Fig.~\ref{fig1}(b)) \cite{PhysRevB.65.195315,PhysRevB.76.045331}. Cancelling this FSS has long been identified as a key requirement for the on-demand generation of polarization-entangled photon pairs via the radiative biexciton-exciton cascade \cite{Benson2000,liu_solidstate2019}. Similarly, near-zero FSS was shown to enable high fidelity initialization of long-lived quantum dot hole spin qubits~\cite{BrashPRB2015}. Conversely, it was very recently shown that the brightness of an exciton based single-photon source under resonant pumping could be optimized for an optimal  finite FSS value~\cite{Ollivier2020}. All these perspectives have led to the development of a large variety of techniques to control the FSS.  Material growth and processing are tailored to reduce the average  FSS, for instance growing small QDs~\cite{stevenson2006}, performing post growth annealing~\cite{YoungPRB2005} or using droplet epitaxy~\cite{huo2013}. Fine tuning of the residual FSS  was explored through the application of both magnetic and electric fields~\cite{Kowalik2005,gerardot_manipulating_2006,stevenson2006} as well as  strain ~\cite{seidl_2006,kuklewicz2012}.
However, the control of a single degree of freedom  only leads to zero FSS if the applied field points along the major axis of QD asymmetry~\cite{gong2011,Plumhof2011,bennett2012}. To universally eliminate the FSS, it was theoretically shown that two independent and non-parallel external controls are needed~\cite{TrottaUniversalRecovery}, e.g. biaxial strain or two vector components of an applied electromagnetic field. 
Adding a third control further allows independent tuning of the FSS and the average exciton transition energy~\cite{trotta_energy-tunable_2015}.\\

\begin{figure*}
    \includegraphics{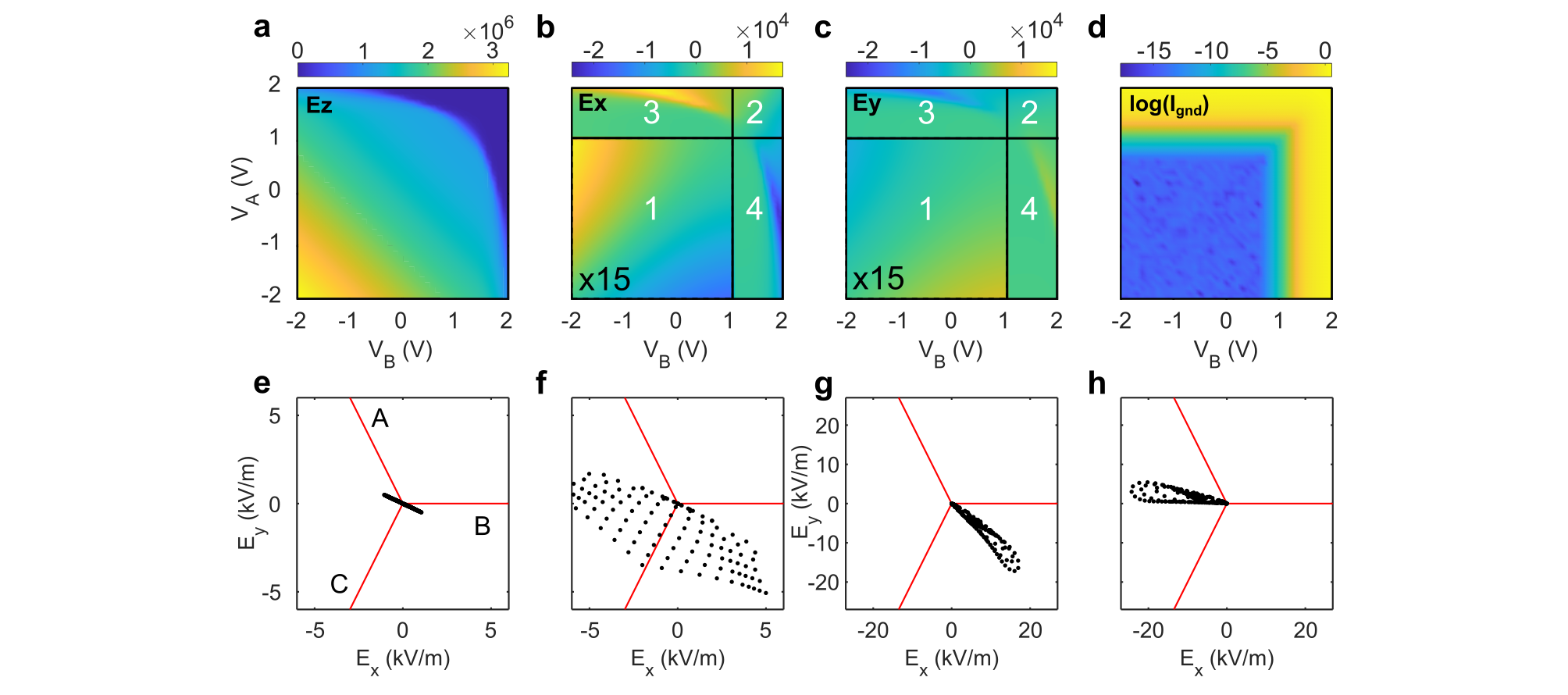}
    \caption{ Finite elements method (FEM)  simulations of the electric field at the cavity center as a function of two applied electric potentials V$_A$ and V$_B$. The third contact is simulated without fixed potential.{ (a-c) Calculated electric field $E_z$  $E_x$ and $E_y$. We define 4 regions labelled 1 to 4 corresponding to (1) non passing regime (2) two-diode passing regime (3,4) one diode passing regime as evidence by (d). The in-plave field values in region (1) has been multipled by 15 for better vizualization with the linear color scale in (b) and (c).  All fields are in units of $V/m$. (d) Calculated net current flowing through the device {in units of $A$, logarithmic scale}.  (e-h) Polar map of the in-plane electric field for the four (V$_A$,V$_B$) voltage regions labelled 1,2, 3 and 4 as indicated in (b),(c). The red solid lines correspond to the orientation of the three narrow ridges.}}
    \label{fig2}
\end{figure*}

Over the years, the toolbox for controlling the excitonic fine structure has become quite diverse and efficient, leading to  record entanglement fidelity for entangled photon pairs~\cite{huber2018}. However, such control has not yet been achieved for QDs inserted in optical cavities,  a crucial step to ensure high collection efficiency of the single and entangled photons\cite{Wang2016,Somaschi2016,dousse_ultrabright_2010}, and for reaching near unity {photon}  indistinguishability~\cite{Somaschi2016,Wang2016,Tomm2021}.
\\

Here, we propose a structure that allows controlling the excitonic FSS for quantum dots inserted in monolithic micropillar cavities. We define three independent control knobs by remotely applying three electric potentials on large p-i-n diodes that are connected to the pillar cavity through narrow ridges. {The vertical p-i-n doping profile extends over the full device.} Finite element simulations are conducted to monitor both the stationary electric field and current in the structure for various bias situations, evidencing a control of the electric field at the position of the QD in all three directions of space. We experimentally explore this approach and {demonstrate both a control of the FFS amplitude and of the orientation of the {exciton transition} dipoles. We find that various combinations of applied voltages  reduce the FSS close or below the radiative linewidth. Furthermore,  similar values of FSS can be obtained for  different excitonic wavelengths.}

The structure proposed to control the excitonic FSS for a QD in a cavity is illustrated in Fig.~\ref{fig1}(a). We start from a planar GaAs/AlGaAs microcavity {grown} on a GaAs substrate. For the sample under investigation, the bottom (top) distributed Bragg reflector (DBR) consists of 34 (16) pairs of $\lambda/4-$thick GaAs/AlGaAs layers. The DBRs surround a $\lambda-$thick GaAs cavity spacer with a layer of annealed InGaAs QDs in its center. Vertically, the multilayer structure defines a p-i-n junction with p(n) doping for the top (bottom) mirror and an intrinsic region within the cavity spacer embedding the QDs. Laterally, the  planar structure is etched to define circular pillar cavities with $10$~\textmu m diameters that are  connected through three $3$~\textmu m wide, $50$~\textmu m long ridges to large mesas where titanium-gold  pads are defined for electrical connection. Each pad can be connected to an independent voltage source with the gold-coated back of the substrate defining a common ground potential. By applying various values for the three biases labelled $V_A$, $V_B$ and $V_C$, one can control the amplitude and direction of the electric field at the center of the pillar cavity in the three directions of space. \\

\begin{figure*}
    \includegraphics{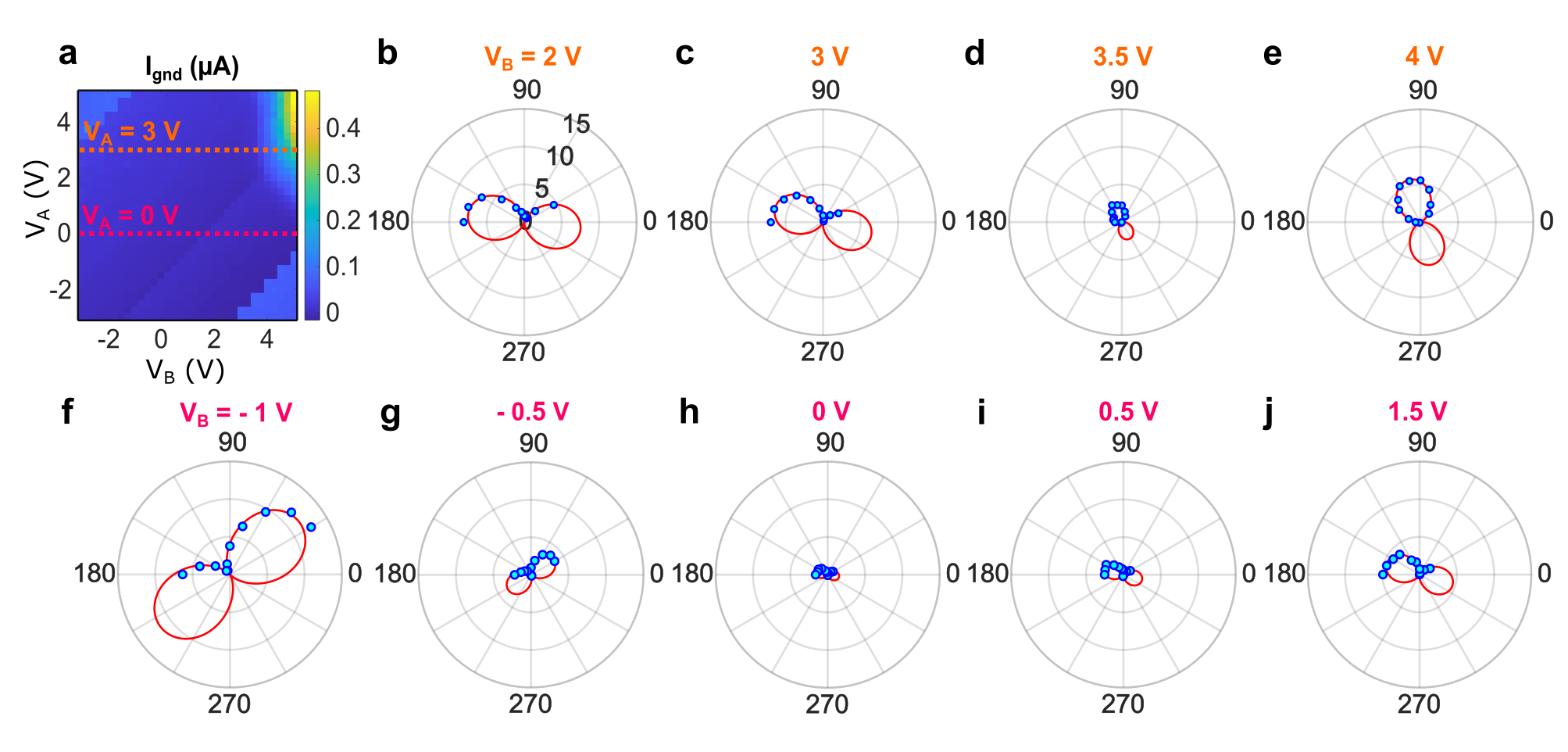}
    \caption{ (a) Measued net current flowing through the p-i-n junction as a function of $V_A$ and $V_B$. (b)-(j) Polar plots of QD1 exciton energy shift as a function of the linear polarization angle of detection (circles) at $V_A=3$~V (b-e) and $V_A=0$~V (f-j) for different voltages $V_B$ as indicated. The solid lines are  fits of the function $\Delta_\text{FSS}( \cos[2(\theta-\theta_0)]+1)/2$. Unit of the radial scale as indicated in (b) is \textmu eV.}
    \label{fig3b}
\end{figure*}

To evidence such a possibility, we first performed  Finite Elements Method (FEM) simulations (COMSOL Multiphysics). For computational efficiency, we replaced the vertical DBR layer structure by a single GaAs slab with a vertical p-i-n doping profile. Such a simplification allows for a qualitative understanding of the physics at play disregarding the complex modulation doping structure that one needs to implement in the {DBR structure}.  Figures~\ref{fig2}(a-c) display the three electric field components at the  center of the micropillar cavity - both in the vertical and horizontal direction -  as a function of two applied electric potentials $V_A$ and $V_B$, while $V_C$ is not fixed by an external potential. 
Along the vertical direction, the calculated electric field $E_z$ presents a large amplitude for negative values of $V_A$ and $V_B$, and progressively decreases when going to the passing regime  for both diodes A and B. The in-plane electric field components, on the other end, show a more subtle behavior and  
 four  regimes can be identified as indicated in the four areas labelled 1 through 4 in Fig ~\ref{fig2}(b-c). They essentially correspond to the regimes of non-passing and  passing p-i-n junction for each of the two electric potentials as evidenced by Fig.~\ref{fig2}(d) presenting the total current flowing through the device (in logarithmic scale).

Figures~\ref{fig2}(e-h) present the {in-plane electric field components} in  polar plots for each regime. For area 1 (Fig.~\ref{fig2}(e)), the amplitude of  the field depends on the applied voltage but its direction remains normal to the unconnected ridge C: the current flows  along the p-doped upper DBR with an in-plane field parallel to it, dictated by the Ohmic resistivity of the p-doped material. 
The direction of the in-plane electric field can be modified when the junctions are in the passing regime. This is the case for area 2 (Fig.~\ref{fig2}(f)) for which {the accessible values of in-plane electric field cover an angular range of almost $\pi$}. Effectively, we generate two independent non-parallel current components which are both drained to the common ground on the back side of the p-i-n junction. Finally, there are two intermediate cases found in areas 3 and 4 of the field maps for which one diode only is in the passing regime. For area 3 (4), the field mostly points in the direction of diode A (B), yet with a noticeable steering effect resulting from the second applied voltage presenting a current drain on the top surface.  

Based on the above discussion, we find that there are at least two ways to generate in-plane electric fields with directional control at the position of a QD centered in the micropillar device. First, in the limit of a non-passing p-i-n junction, using the third control knob $V_C$ (in addition to $V_A$ and $V_B$) would allow to choose the field amplitude in the three directions that are normal to each one of the ridges as it was the case in the direction normal to ridge C with only one pair of voltages ($V_A$ and $V_B$) in Fig.~\ref{fig2}(e).
 Second, in the limit of large forward bias, two electric potentials suffice to control both orientation and magnitude of the field. The maximum attainable magnitude of the in-plane electric field differs by roughly a factor of four between the regimes.

\noindent The link between the observed in-plane electric field behavior and current flows is shown in supplementary figure 2. In all regimes, except for area 2, the current and electric field are roughly parallel. For area 2, two  non-parallel current components are created which are both drained through the p-i-n junction in the passing regime. We underline however that the absolute value of the calculated current is largely over-estimated numerically, because of the strongly simplified doping structured considered in the z direction, as experimentally evidenced hereafter.

\begin{figure*}
    \includegraphics[width=12cm]{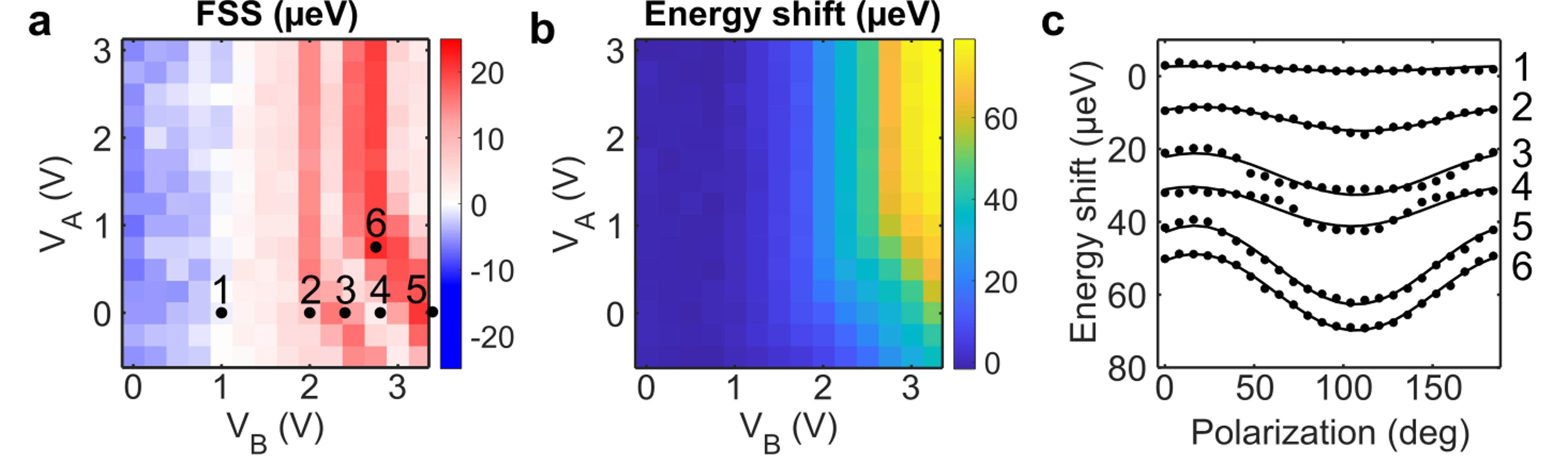}
    \caption{Measurements for QD2. (a-c) Color maps as a function of $V_A$ and $V_B$ of (a) algebraic $\Delta_\text{FSS}$ measured in a fixed basis (see text), (b)  energy shift of the average exciton  transitions with respect to energy at $V_A=V_B=0$.  (c) Energy shift  as a function of detection polarization angles for the 6 points indicated in (a).}
    \label{fig4}
\end{figure*}

We tested this FFS control scheme experimentally on 10~\textmu m diameter pillar cavities. Such large diameter cavities do not show discrete optical modes, so that any cavity bi-refringence effect that could perturb the polarization analysis of the excitonic fine structure is avoided~\cite{strainpillar}.
The  FSS is optically probed through polarization-resolved micro-photoluminescence (µPL) studies performed at 4K, under quasi-resonant (p-shell) continuous-wave excitation \cite{Gazzano2013}. We present experimental data for two QDs (QD1 and QD2) inserted in two distinct cavities. Figure~\ref{fig3b} shows the data set obtained for QD1. Figure \ref{fig3b}(a) presents the measured sum of currents {flowing through the connections to} the two independent voltage sources. This {sum corresponds the portion of the current that flows through the p-i-n junction, since the horizontal current flowing on top of the device cancels out}.  Figure~\ref{fig3b}(a) allows one to identify the various regimes for the FSS control {and evidences maximal current in the $\mu A$ range, orders of magnitudes lower than predicted by our simplified  numerical simulations}. The passing regime is obtained for $V_A \ge 2~V$ and $V_B \ge 4~V$.  
 This asymmetry, that is not accounted for in our numerical simulations, can be due to the imperfections of the 1D ridges  exhibiting different {Ohmic} resistances.\\ 
 
 The small FSS of our QDs is not spectrally resolved by our setup, however the exciton  photoluminescence peak energy undergoes a clear sinusoidal shift of amplitude $\Delta_\text{FSS}$ in the 10~\textmu eV range as a function of the angle $\theta$ of the detected linear polarization. This measurement is achieved    here by rotating  a half-wave-plate  before a fixed polarizer ({See supplementary}). In Fig.~\ref{fig3b}(b-e),   polar plots  of this shift are reported  for  different voltages $V_B$ at a fixed voltage $V_A=3$~V. For each voltage, the  angular dependence is fitted by the function $\Delta_\text{FSS}( \cos[2(\theta-\theta_0)]+1)/2$ where $\theta_0$ refers to the polarization angle of the high energy exciton line. A reduction of   $\Delta_\text{FSS}$ down to {$ 2.4\pm 0.6 $~\textmu eV} is observed at $V_B=$3.5~V, at the edge of the passing regime,  together with a rotation of the exciton eigenaxes on the order  of $\pi/2$. This is the characteristic signature of near cancellation of the FSS in such a way that the eigenaxes of the high and low energy exciton lines are exchanged when (ideally) crossing the zero FSS point~\cite{bennett2012,TrottaUniversalRecovery}.
 A further reduction {down to $1.55\pm 0.20$~\textmu eV} is observed  at a fixed voltage $V_A=0$~V (non-passing regime) when varying $V_B$, as seen in Fig.~\ref{fig3b}(f-j). For both the  passing   and non-passing regimes, a  significant reduction of the FSS near or below the radiative linewidth is thus demonstrated, whereas over the whole voltage scans, we  observed significant variations of  FSS amplitude up to $\sim 20$~\textmu eV.

To complete the experimental picture, we report similar results for QD2 in Figure~\ref{fig4}. {Here,  the exciton peak energy is measured for each value of ($V_A,V_B$) for  the exciton eigenaxes identified for $V_A=V_B=0$~V. We deduce an \emph{algebraic} value for FSS defined as the energy difference between these two energies. } Figure~\ref{fig4}(a) shows that the  algebraic FSS  varies over more than 20~\textmu eV and exhibits a clear  transition  at the voltage  $V_B\approx 1$~V where the algebraic FSS changes sign. To  determine whether it is due to FSS cancellation (and not only to a rotation of the QD eigenaxes~\cite{bennett2012,TrottaUniversalRecovery}), the exciton peak energy is measured as a function of the angle of detected polarization, for 6 different  points as indicated in Fig.~\ref{fig4}(a). The sinusoidal dependence  reported in Fig.~\ref{fig4}(d) does not show any significant rotation of the eigenaxes in the probed region of positive FSS and a vanishing FSS of $ 1.3\pm 0.7$~\textmu eV, below the radiative linewidth for point 1 in Fig.~\ref{fig4}(a). Besides, when considering a fixed value of $V_A=0\  V$ and increasing $V_B$, the FSS turns out to vary non  monotonously, showing  notably  a local minimum at $V_B=2.75$~V, a  behavior also observed  for QD1. Such an observation is not directly explained by the electrical field which, according to our numerical simulations, should evolve monotonously. This rather reveals the complex dependence of  FSS on the electric field, as already reported in Ref.~\onlinecite{gerardot_manipulating_2006} for an in-plane application of an electric field. Combined with the average exciton energy shift due to the quantum confined Stark effect, covering a range around 80~\textmu eV as shown in Fig.~\ref{fig4}(b), this behaviour is of particular interest to reduce or modify  the FSS for different central exciton energies. Such a possibility is evidenced in Fig.~\ref{fig4}(b) where similar FSS values are obtained at different exciton energy {for voltage combination} 3 and 4 as well as for 5 and 6.

\noindent {Finally, we underline that the  method proposed to control the exciton FSS relies on a geometry very similar to the one already used to obtain highly indistinguishable photons under resonant excitation. This was reproducibly demonstrated from QDs deterministically positioned in micropillar cavities controlled by a single electrical pad \cite{Somaschi2016,Ollivier2020}, in a large applied voltage range with limited  current. We thus anticipate that the efficient and versatile control of the FSS obtained {here} in the non-passing regime would allow high indistinguishability. We note, however, that significant emission linewidth broadening is observed here in the passing regime ($V_B>3$~V) in the present time-integrated measurements obtained under non-resonant excitation (see supplementary figure S3). }

In conclusion, we have proposed a new scheme  to control the exciton fine structure of a QD inside a micro-pillar cavity based on three independent electric potentials. 
Numerical simulations provide important insight on how the proposed geometry allows for a three-dimensional control of the electrical field at the center of the pillar cavity. Experimentally, we observe a tuning  of  the FSS in the 10-20~\textmu eV range   \cite{trotta2014,trotta_energy-tunable_2015} which is shown to be sufficient  to cancel the FSS of annealed InGaAs QDs. {The  mean-exciton energy is controlled over a   significantly lower range than the one obtained for strain tuning techniques in bulk samples. 
However,  several parameters  can be explored to increase the amplitude of the electric field variation in the pillar center: length and width of the connecting ridges, vertical doping structures, etc. Here, we reported an 80~\textmu eV exciton tuning range - a value that is already} sufficient to tune the QD exciton  into resonance with the pillar cavity mode when combined with the in-situ lithography technique \cite{dousse_controlled_2008}. Indeed, this technique has been shown powerful to define a quasi-resonant pillar cavity around a chosen quantum dot for bright single photon sources or entangled photon pairs~\cite{Somaschi2016,dousse_ultrabright_2010}. Our work thus opens the path toward the  fabrication of bright sources of single photons and entangled photon pairs, with control of both the QD-cavity detuning and the excitonic fine structure.

\vspace{0.5cm}
\textbf{Acknowledgements} 

This work was partially supported by the H2020-FET OPEN project number 899544 - PHOQUSING, the French RENATECH network, and the European Research Council Starting Grant No. 715939, Nanophennec. H. O. acknowledges support from the Paris Ile-de-France Région in the framework of DIM SIRTEQ. M.E. acknowledges funding by the Deutsche Forschungsgemeinschaft (DFG, German Research Foundation) Project 401390650, and by the University of Oldenburg through a Carl von Ossietzky Young Researchers fellowship.

\bibliography{FSSpaper_arxiv}


\clearpage
\pagebreak
\widetext
\begin{center}
\textbf{\large Supplementary Information:\linebreak}

\textbf{\large Three-dimensional electrical control of the excitonic fine structure  for a quantum dot in a  cavity}
\end{center}

\vspace{1cm}

\setcounter{figure}{0}
\makeatletter
\renewcommand{\thefigure}{S\arabic{figure}}

\begin{figure}[ht!]
    \includegraphics{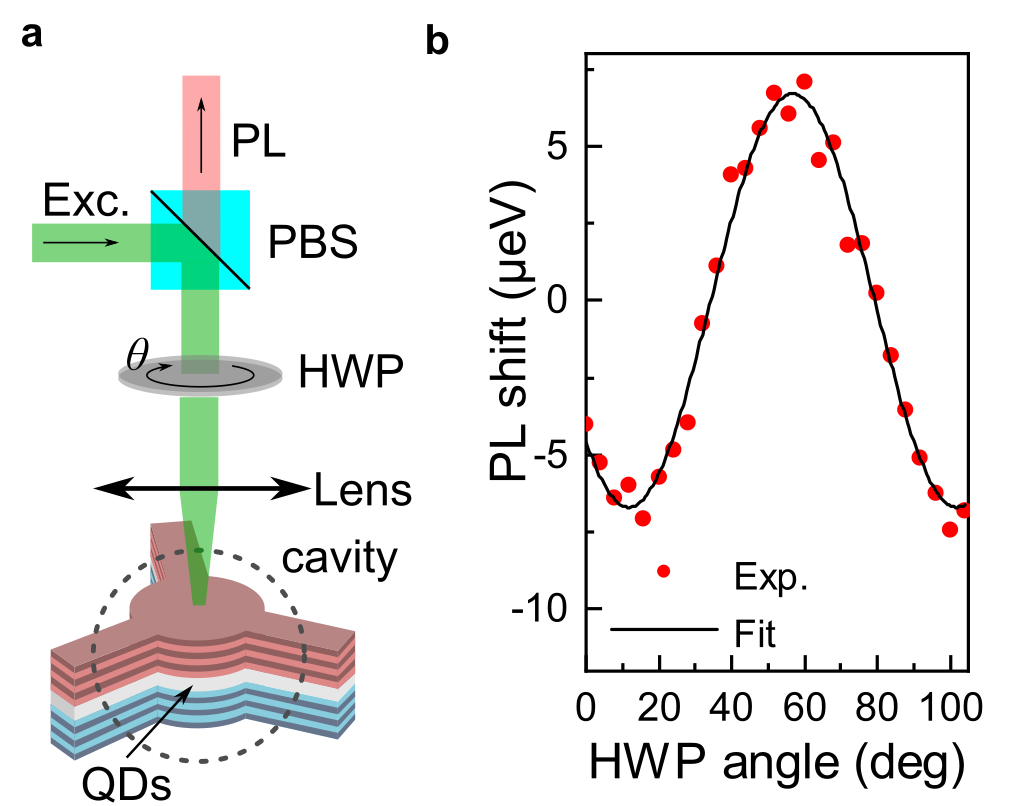}
    \caption{\textbf{Experimental procedure}. \textbf{a} Scheme of polarization-resolved micro-photoluminescence (µPL) measurement in quasi-resonant (p-shell) excitation. We optically probe the excitonic FSS  by polarization-resolved micro-photoluminescence (µPL) under quasi-resonant (p-shell) excitation.  Optical excitation and collection are realized through the micropillar top surface and the combination of  half-wave plate (HWP) and a polarizing beam splitter (PBS) allows accessing polarization-dependence of the collected photoluminescence (PL).  \textbf{b} Example data set obtained for QD2 showing the spectral position of the exciton line which exhibits a sinusoidal shift  as a function of HWP angle corresponding to  $\Delta_{FSS}\approx10\mu eV$. The extremes of the sine represent positions where the collection is oriented along one of the excitonic eigenaxes. Note that for the annealed QDs under study, the FSS is smaller than our spectral resolution.}
    \label{figSI_1}
\end{figure}

\begin{figure}
    \includegraphics{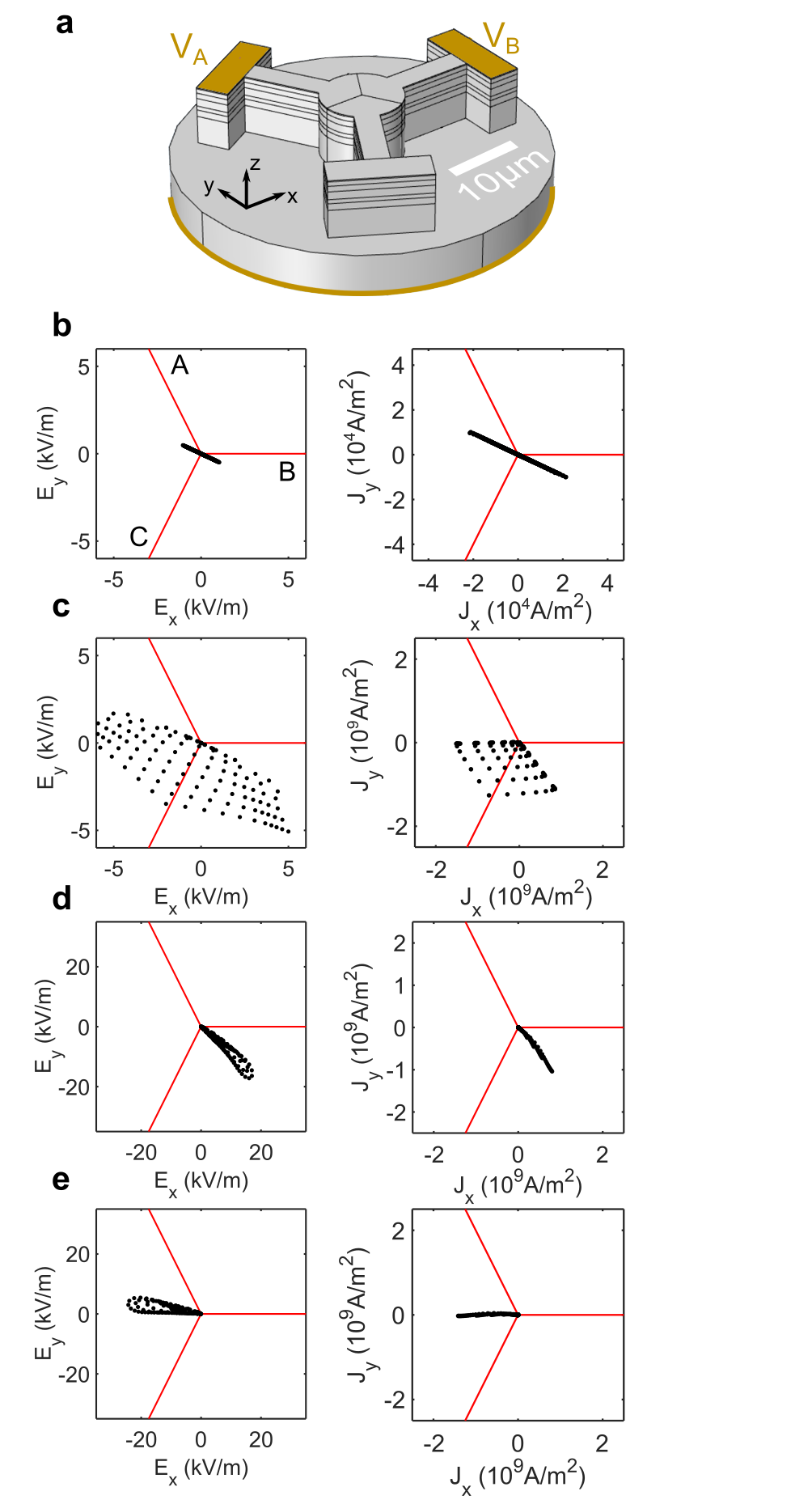}
    \caption{\textbf{Simulated current and electric field}  \textbf{a} Schematic of the structure modelled in the Finite Elements Method (FEM) simulations (COMSOL Multiphysics) \textbf{b-e} Polar plots of the in-plane electric field  and corresponding in-plane current for the voltage regions labelled  1 to 4 in the main text Figure 3b and 3c.}
    \label{figSI_2}
\end{figure}

\begin{figure}
    \includegraphics[width=6cm]{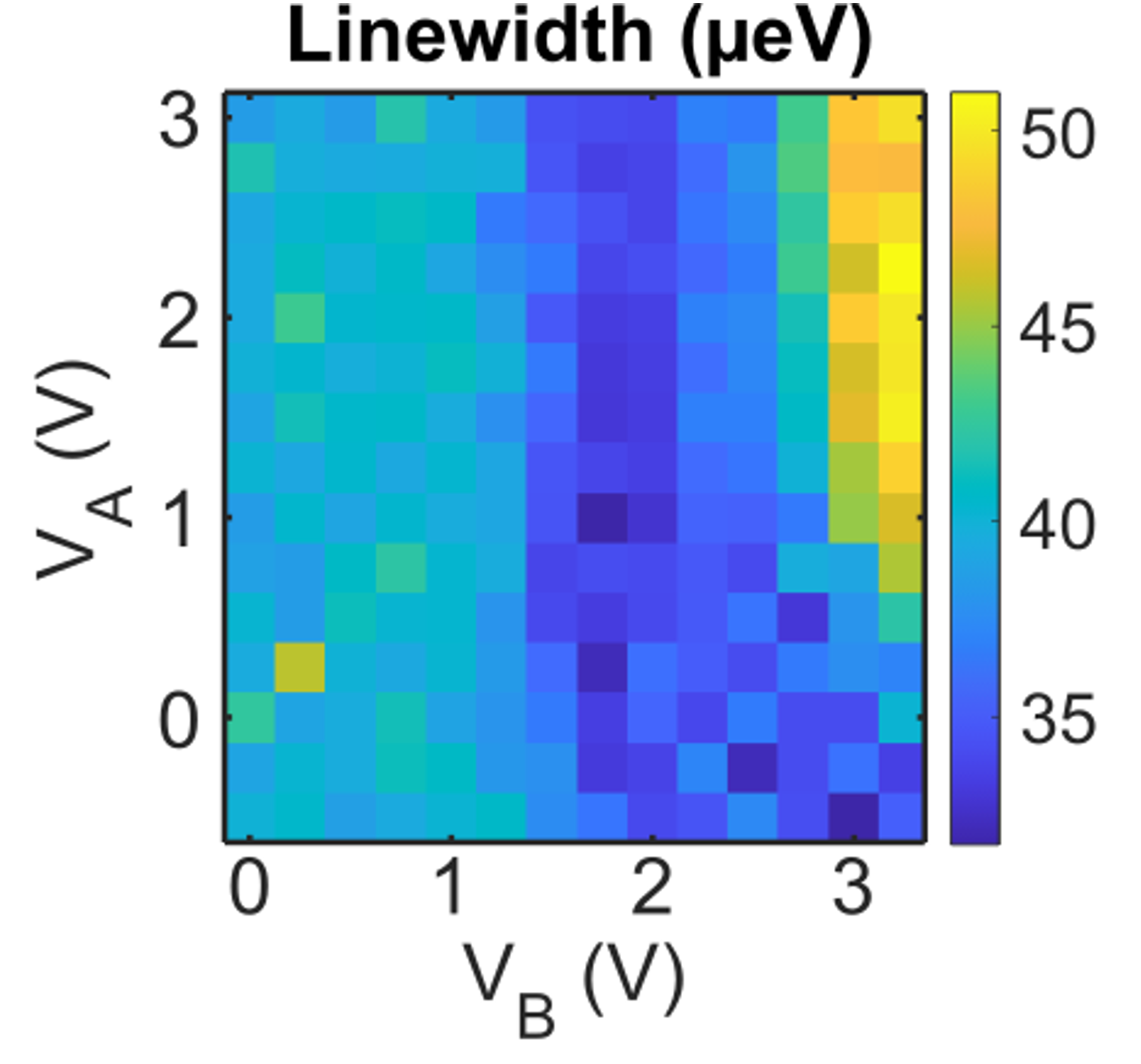}
    \caption{\textbf{QD2 linewidth} Color map of the average linewidth measured for QD2 as a function of $V_A$ and $V_B$. The observed values, larger than  our apparatus spectral resolution of about {23.4~\textmu eV  (confirmed)} reveal a significant broadening  mostly due to the charge noise induced by the non-resonant excitation. Such a broadening would disappear under resonant excitation as demonstrated in \cite{Somaschi2016,Ollivier2020}. However, an additional broadening is  observed in the passing regime ($V_B>3$~V, $V_A>1$~V), most probably induced by the current flowing through the structure in the passing regime.}
    \label{figSI_3}
\end{figure}

\end{document}